\documentclass[a4paper,11pt]{article}
\usepackage{jheppub} 

\usepackage[T1]{fontenc} 

\usepackage{hyperref}
 \usepackage{amsmath}
\usepackage{amssymb}
\usepackage{epsfig}
\usepackage{natbib}
\usepackage{epstopdf}
\usepackage{graphicx}
 \usepackage{csquotes}

\newcommand*{\be}{\begin{equation}}
\newcommand*{\ee}{\end{equation}}
\newcommand*{\bea}{\begin{eqnarray}}
\newcommand*{\eea}{\end{eqnarray}}

\title{\boldmath   The energy-frequency diagram of  the (1+1)-dimensional $\Phi^4$ oscillon }

\author[a]{N. V. Alexeeva,}
\author[b,1]{I. V. Barashenkov, \note{Corresponding author.}}
\author[a,c]{Alain Dika,}
\author[a]{and Raphael De Sousa}

\affiliation[a]{Department of Mathematics and Applied Mathematics, University of Cape Town, Rondebosch 7701, South Africa}
\affiliation[b]{Centre for Theoretical and Mathematical Physics, University of Cape Town,  South Africa}
\affiliation[c]{Joint Institute for Nuclear Research, Dubna 141980, Russia}

\emailAdd{Nora.Alexeeva@uct.ac.za}
\emailAdd{Igor.Barashenkov@uct.ac.za}
\emailAdd{A.Dika@uct.ac.za}
\emailAdd{dssrap001@myuct.ac.za}

\abstract{
Two different methods are used to study the existence and stability of the (1+1)-dimensional $\Phi^4$ oscillon.
The variational technique approximates it by a periodic function with a set of adiabatically changing parameters.
An alternative approach treats oscillons as  standing waves in a finite-size box;
these are sought as solutions of a boundary-value problem on a two-dimensional domain.
The numerical analysis reveals that the standing wave's 
energy-frequency diagram is fragmented into disjoint segments with
$\omega_{n+1} < \omega < \omega_{n}$,  where $\omega_n= \omega_0/ (n+1)$, $n=0,1,2, ...$,
and $\omega_0$ is the endpoint of the continuous spectrum (mass threshold of the model).
The variational approximation involving the first, zeroth and second harmonic components provides an 
accurate description of the oscillon with the frequency in $(\omega_1, \omega_0)$, but breaks down
as $\omega$ falls out of that interval.
}

\begin{document} 
\maketitle
\flushbottom

\section{Introduction}

Oscillons    (also known as pulsons)
were introduced as localised long-lived  pulsating
 structures in three-dimensional classical field theories    \cite{Voronov1,BM1,BM2,G1,CGM}. 
 The original motivation  \cite{Zeldovich}
 was to model the vacuum domain formation in theories with spontaneous symmetry breaking 
 and its cosmological implications.  
 Oscillons have now been recognised to have a role in
 the dynamics of inflationary reheating,  symmetry-breaking phase transitions, and  false vacuum decay
  \cite{CGM, Riotto, GInt,  11cosmo,  Dymnikova,  Broadhead,   bubbling,  Amin1,  Stamatopoulos,  Zhou,  Amin2,   Adshead,   GG,Bond,   Antusch,   Hong,  Cyn, LozAm, Takhistov, Aurek}.
  They arise as natural ingredients in
   the
   bosonic sector of the standard model 
 \cite{Farhi2,  Graham, Gleiser4, Sfakianakis, Piani}
 and
   axion-based models of 
   fuzzy dark matter  \cite{Kolb,Vaquero,Kawa_axion,Olle,Arvanitaki,Miyazaki, string,Kasu,Sang,Imagawa}.
  The Einstein-Klein-Gordon equations have also been shown to exhibit oscillon solutions
  \cite{Maslov,Zhang2,Nazari,Kou1,Hira,Kou2,Naka}.
 % axion models \cite{Kolb,Vaquero,Kawa_axion,Olle,Arvanitaki,Miyazaki},  string phenomenology \cite{string,Kasu,Sang}  
% The (2+1)-dimensional oscillons have been studied     in  the context of the planar Abelian Higgs theory \cite{GT2,Achi}.

For some time, the studies of three-dimensional  oscillons were disconnected from research into oscillatory solitons in 1D.  
The reason was that the latter objects --- interpreted as the kink-antikink bound states   \cite{Kudryavtsev, DHN,Getmanov,Sugiyama,Geicke,CP} ---  were believed to persist over long periods 
of time (or even indefinitely),
 emitting little or no radiation.
By contrast, the three-dimensional oscillons were thought to have a fairly short lifespan  \cite{BM1,BM2,G1,CGM}.

More  accurate mathematical and numerical analysis indicated, however,  that the oscillons in three dimensions   and 
one-dimensional bound states  
 share their basic properties
 \cite{Honda,Fodor1,Fodor3}.   (The only exception, of course, is the sine-Gordon breather --- an exactly periodic solution  which emits strictly no radiation.) 
It is for this reason that we are using the {\it oscillon\/} nomenclature for what would otherwise be called    \enquote{bion}    \cite{Getmanov,BK},     \enquote{approximate breather} 
 \cite{SK} or  \enquote{breather-like state  % Include Cambell-Wingate?
 on the line}   \cite{Vachaspati}.

In this paper, we employ a  new variational approach
to
study localised oscillations in the one-dimensional 
$\Phi^4$   equation  --- the one-dimensional version of the system that bore the originally discovered pulsed states  \cite{DHN,Kudryavtsev,Voronov1,BM1,BM2}. 
%The analysis of this model with spontaneously broken symmetry proves to be more complicated than  the variational treatment of   the model with a symmetric vacuum   \cite{BA}. 
 We consider our present attack on the one-dimensional $\Phi^4$ oscillon  a step towards the consistent variational description 
 of its three-dimensional counterpart.

The 1+1 dimensional $\Phi^4$ theory  is probably the simplest model with spontaneous symmetry breaking exploited in the studies of quantum fields
\cite{Rychkov,Bajnok,Serone,Bordag,Graham-Weigel,Tokacs,Martin,Ito} and  phase transitions \cite{PT1,PT2}.
The model is 
defined by the Lagrangian
\be
L=    \frac12   \int  \left[ \Phi_t^2- \Phi_x^2  - (\Phi^2-1)^2   \right]  \, dx
\label{L0}
\ee
and equation of motion 
 \be
\Phi_{tt} - \Phi _{xx}
- 2 \Phi (1-\Phi^2) =0.
\label{A0}
\ee
Its oscillon solution was discovered  by Dashen, Hasslacher and Neveu \cite{DHN} who 
 constructed it
 as an asymptotic series in powers of the amplitude of the 
oscillation. The first few terms in  the  expansion of \cite{DHN}
 (with typos corrected) are
\begin{align} 
\Phi= 1  + \frac{\epsilon}{\sqrt{3}}   \,  \mathrm{sech} (\epsilon x) \cos (\omega t)  
+     \frac{\epsilon^2}{12}  \mathrm{sech}^2 (  \epsilon x) 
 \left[
\cos (2 \omega t) -3 
\right]  \nonumber \\
+ \frac{\epsilon^3}{144  \sqrt 3} [ 98 \, \mathrm{sech} (\epsilon x) - 103 \, \mathrm{sech^3}(\epsilon x) ] \cos(\omega t)  
+ \frac{\epsilon^3}{48 \sqrt 3} \mathrm{sech^3} (\epsilon x) \cos ( 3 \omega t)+ O(\epsilon^4).
\label{A1}
\end{align} 
Here $\omega^2 = \omega_0^2-  \epsilon^2$,    $\epsilon \to 0$
and $\omega_0$ denotes the mass of elementary constituents of the model
(the endpoint of the  continuous spectrum   of  vacuum
excitations):
\[
\omega_0=2.
\]

  Segur and Kruskal \cite{SK} proved that the series \eqref{A1} does not converge 
  and that  the true oscillon expansion includes terms that are nonanalytic in $\epsilon$. 
  (For the system-dynamic perspective on this argument, see Eleonskii  {\it et al \/}  \cite{Eleonskii}.) 
  The nonanalytic terms lie beyond all orders of the perturbation theory and make negligible contributions to the core of the oscillon;
  yet they do not vanish as $|x| \to \infty$ and account for the oscillon-emitted radiation.
  Boyd \cite{Boyd2} noted a close relationship between radiating (hence  decaying) oscillons of small amplitude
  and {\it nanopterons:} standing waves with    \enquote{wings} extending to the infinities. (See also  \cite{Fodor2}.) 
  The amplitude of the wings is exponentially small in $\epsilon$ but does not decrease as $x \to \pm \infty$. 
The sum of the standing wave and a solution to the linearised equation with an exponentially small amplitude gives a highly accurate approximation of the radiating oscillon \cite{Boyd2}. 
   
 Regarding oscillons of finite amplitude, numerical simulations have been the primary source of information.
 The earliest observations of oscillons with finite $\epsilon$ are due to Kudryavtsev \cite{Kudryavtsev}, while more detailed and accurate sets of simulations can be found in \cite{Getmanov,Geicke,Boyd2,Fodor2}. 
For reviews, see \cite{BK,Fodor3,Boyd1}.  
 
 The present study is motivated by the need
to have an analytic  tool capable of providing insights into the structure and properties  of the finite-amplitude oscillons,
 similar to the collective coordinate technique used in the nonlinear Schr\"odinger domain \cite{Malomed,BAZ}. 
  Modelling on the 
 multiscale variational method developed for the theory with the symmetric vacuum \cite{BA},
 we formulate a variational approach to the one-dimensional $\Phi^4$ oscillon. 
The  symmetry-breaking  nature of the vacuum in \eqref{L0} forces us to
 expand the set of collective coordinates that was employed in \cite{BA}.  However, similar to \cite{BA},
  the expanded set does not include any radiation degrees of freedom.  We approximate the oscillon by a strictly periodic, nonradiating,  state.
  
  To validate our variational approximation, we carry out a numerical study of  the 
  time-periodic solutions of equation \eqref{A0} on 
  a finite interval. 
  The earlier studies determined  standing waves with frequencies $\omega > \omega_0 /  \sqrt 2$ \cite{Boyd1}; we will reach below that record.

  For each $\omega$, there is a continuous family of standing waves with different amplitudes of the radiation wings and, consequently, different energies.
   %The energy of the standing wave with a given frequency is sensitive to the interval length. 
   We focus on the waves with the lowest energy as these  nanopterons have the
    smallest wing amplitude   \cite{Fodor2,Fodor3,Fodor1}. 
  The energy-frequency diagram of the lowest-energy standing waves 
 is found to be in good agreement with the diagram constructed variationally in a wide range of frequencies.

 The paper is organised into five sections. The variational approximation for the $\Phi^4$ oscillons 
 is presented in Section \ref{multi}, with some technical details  relegated to the Appendices. 
 In Section \ref{numerical}, we determine numerical solutions describing standing waves in a finite-size box,
 and in Section \ref{comparison}, the  results of the variational and numerical approaches are compared. 
 % Section \ref{stability} deals with the stability of the fixed-point solutions of the variational equations. 
 Section \ref{conclusions} summarises conclusions of this study.

\section{Multiscale variational approach}
\label{multi}
\subsection{Method} 

The variational method is arguably the main analytical approach to solitons in nonintegrable systems
outside the perturbation expansions. In the context of oscillons, the method was pioneered in
 Ref \cite{CGM}, where  the three-dimensional $\Phi^4$ oscillon  was  approximated 
 by a localised waveform 
\be
\Phi= 1+A(t) e^{-(r/b)^2}.
\label{B1}  \ee
Here $A(t)$ is an unknown oscillatory function describing the trajectory of the structure's central point 
and $b$ is an arbitrarily chosen value of its width.  (Ref \cite{Kev} followed a 
similar strategy when dealing with the two-dimensional sine-Gordon equation.) Once  the ansatz \eqref{B1} has been substituted  in the lagrangian and  the $r$-dependence 
integrated away, the variation
 of action produces 
a second-order equation for  $A(t)$. 

The regular variational  method does not suggest any optimisation strategies for the choice of $b$. 
Making $b(t)$ another collective coordinate  --- as it is done in the studies of the nonlinear Schr\"odinger solitons \cite{Malomed,BAZ} ---
gives rise to an ill-posed dynamical system  not amenable to numerical simulations 
 \cite{CF,BA}. This difficulty is circumvented in the multiscale approach,   which considers the oscillon as a rapidly oscillating structure with
 adiabatically changing parameters \cite{BA}.

Following \cite{BA},  we consider 
  $\Phi$  to be a function of two  time variables, ${\mathcal T_0}=t$ and 
  ${\mathcal T_1}= \epsilon t$.
    The rate of change  is assumed to be $O(1)$ on either scale:
   $\partial \Phi/ \partial \mathcal T_0, \partial \Phi/ \partial \mathcal T_1 \sim 1$. 
   We require  $\Phi$ to be periodic in $\mathcal T_0$, with a period of $T$:
   \[
   \Phi(\mathcal T_0+ T; \,   \mathcal T_1)=\Phi(\mathcal T_0; \,  \mathcal T_1).
   \]
   
As $\epsilon \to 0$, the variables $\mathcal T_0 $ and $\mathcal T_1$  become independent and 
   the Lagrangian \eqref{L0} transforms into 
   \begin{align} 
L= \frac12
 \int \left[   \left( \frac{\partial  \phi}{\partial {\mathcal T_0} } + \epsilon \frac{\partial  \phi}{\partial {\mathcal T_1} }   \right)^2
-\phi_x^2   
-4 \phi^2    -4\phi^3 -\phi^4 
 \right] dx,
 \label{A2}
\end{align} 
where we let $\Phi=1+\phi$.
    The action $\int L dt$ is replaced with
\be
\label{A3}
S= \int_0^T  d   {\mathcal T_0}   \int d   {\mathcal T_1} \,  L   \left(    \phi,  \frac{\partial  \phi}{\partial {\mathcal T_0} }, 
   \frac{\partial  \phi}{\partial {\mathcal T_1}}     \right).
\ee
Modelling on the  asymptotic expansion \eqref{A1}, we choose the trial function 
in the form
   \begin{align}
\phi= A \,  \cos (\omega {\mathcal T_0} +\theta ) \, \mathrm{sech}  \left(   \frac{x}{b}   \right)    
+ \left[  B + C   \cos 2( \omega {\mathcal T_0} +\theta )   \right]  \, \mathrm{sech}^2  \left(   \frac{x}{b}   \right),  
 \label{A4}
\end{align}
where $A, B, C,  b$ and $\theta$ are functions of the   \enquote{slow} time variable $\mathcal T_1$
while $\omega = 2\pi/T$ ($\omega>0$).

Once the explicit dependence on $x$ and $\mathcal T_0$ has been integrated away, 
equations \eqref{A2} and \eqref{A3} give
\be
S=  T \int d  {\mathcal T_1}  \, \mathcal L_\omega,
 \label{A5}
\ee
where the effective Lagrangian $\mathcal L_\omega$ is given by 
\be
\mathcal L_\omega = K+  \frac{b}{2}    \left(A^{2}+\frac{8 C^{2}}{3}\right)       \left( \dot \theta+ \omega\right)^2-U, 
\label{Lw}
 \ee
with
\begin{align}
K=          \frac12  \dot A           (bA     \dot ) + \frac23             \dot B             (bB  \dot )+ \frac13            \dot C      (bC \dot ) 
%  \nonumber\\
+ \frac{  {\dot b}^2}{2b} 
 \left[ \left( \frac{\pi^2}{36}  + \frac13 \right)   A^2 + \frac{2 \pi^2}{45} (2 B^2 +  C^2) \right]
%   +\left(A^{2}+\frac{8 E^{2}}{3}\right) b      \left( D\theta+ \omega\right)^2
\label{A100}
  \end{align} 
  and 
    \begin{align}
  U=     \left(  \frac{1}{6b}    +2b \right)A^2 +   \frac{4}{15}     \left( \frac{1}{b}   +5b \right) \left( 2B^2+C^2    \right)              \nonumber  \\
  +2b  \left[  \frac{8}{15} B (2B^2+3C^2) + A^2 (2B+C) \right]     \nonumber   \\
  +  b  \left[
  \frac{A^4}{4} + \frac{16}{35}  \left( B^4 +\frac38 C^4 + 3B^2C^2 \right) 
  + \frac{4}{5} A^2  \left( 2B^2+C^2+ 2BC \right)
  \right]. 
 \label{A6}
  \end{align}
  In \eqref{A100}, we introduced a short-hand notation
  $\dot f \equiv \epsilon \partial f/  \partial  \mathcal T_1$. 
  
The variation of the lagrangian \eqref{Lw} with respect to the collective 
coordinates $A, B, C, b$ and $\theta$ produces five equations of motion. The variable $\theta$ is cyclic; the corresponding Euler-Lagrange equation 
gives rise to the conservation law
\be
b   \left(A^2+\frac83 C^2 \right) ( \dot \theta + \omega) = \ell,
 \label{AA7}
\ee
where $\ell= \mathrm{const}$.
Making use of \eqref{AA7}, $\dot \theta$ can be eliminated from the remaining four equations which become a system of four equations for 
four unknowns ($A,B, C$ and $b$): 
\begin{subequations}
 \label{vareq}
\begin{align}
	  \ddot{A}+ \frac{ \ddot{b} }{2b} A     +   \frac{\dot b}{b}   \dot{A}
	   - \left( \frac{\pi^2}{36}+\frac13 \right)  \left(   \frac{\dot b}{b} \right)^2 A 
    + \left(4+  \frac{1}{3b^2} \right)A 
    - \frac{9 \ell^{2} }{\left(3 A^{2}+8 C^{2}\right)^{2} }  \frac{A}{b^2} 	
   \nonumber   \\   
+ 4A(2B+C) + A^{3}  
	 + \frac{8}{5} A( 2B^2+ 2BC+ C^2)=0,
	  \label{var-a}    \\    \nonumber \\
	    %%%%%%%%%%%%%%%%%%%%%%%%%%%%%%%%%%%%%%%%%%%%%%%%%%%%%%%%%%%%%%%%%%%%%%%%%%%%%%%%%%%%
	 \ddot B +  \frac{\ddot b}{2b} B +  \frac{\dot b}{b} \dot B 
	 - \frac{\pi^2}{15} \left( \frac{\dot b}{b} \right)^2 B 
	+  \frac{4}{5} \left( 5+ \frac{1}{b^2} \right) B  \nonumber \\
	+ \frac35 (5A^2+8B^2+4C^2)  + \frac{6}{5} A^2 (2B+C) 
	+\frac{24}{35} B (2B^2+3C^2)=0,
	    \label{var-b}   \\      \nonumber \\	  
	    	  %%%%%%%%%%%%%%%%%%%%%%%%%%%%%%%%%%%%%%%%%%%%%%%%%%%%%%%%%%%%%%%%%%%%%%%%%%%%%%%%%%%%
		 	 \ddot C+  \frac{\ddot b}{2b} C +  \frac{\dot b}{b} \dot C
	- \frac{ \pi^2}{15} \left( \frac{\dot b}{b} \right)^2 C+\frac{4}{5} \left( 5+ \frac{1}{b^2} \right) C
	- \frac{36 \ell^2}{(3A^2+8C^2)^2 }\frac{C}{b^2}	
	\nonumber \\
 + 3 \left( A^2 +    \frac{16}{5}  BC \right) 
	+ \frac{12}{5} A^2(B+C) + \frac{36}{35} (4B^2+C^2)C=0,
    \label{var-c} 
	\\	\nonumber  \\	    
	%%%%%%%%%%%%%%%%%%%%%%%%%%%%%%%%%%%%%%%%%%%%%%%%%%%%%%%%%%%%%%%%%%%%%%%%%%%%%%%%%%%%	 	
	 A \ddot A+ \frac43  B \ddot B + \frac23 C \ddot C +
	 \left[ 2 \frac{\ddot b}{b}- \left( \frac{\dot b}{b} \right)^2 \right] \left[ \left( \frac{\pi^2}{36}  + \frac13 \right) A^2 +\frac{2 \pi^2}{45}   (2B^2+ C^2) \right]  \nonumber \\
	 + 4 \frac{\dot b}{b} \left[ \left(\frac{\pi^2}{36} +\frac13 \right) A \dot A + \frac{2 \pi^2}{45} ( 2B  \dot B  + C \dot C) \right] 	 - \frac{3}{3A^2+8C^2} \frac{\ell^2}{b^2}
	  \nonumber \\
	 +\left( 4- \frac{1}{3b^2} \right) A^2 + \frac{8}{15} \left( 5-\frac{1}{b^2} \right) (2B^2+C^2)
	 +4A^2(2B+C) +\frac{32}{15}  B (2B^2+ 3C^2)  \nonumber \\
	 +\frac12 A^4 +\frac{32}{35} \left( B^4+\frac38 C^4 + 3B^2 C^2 \right) +\frac85A^2 (2B^2+C^2+2B C)=0.
	 	 \label{Ap4}
\end{align}
\end{subequations}

 The system of four equations has a Lagrangian
 \begin{subequations}
 \label{Lagell}
 \be
 \label{LL}
  \mathcal L_\ell   = K  - U_\ell,
 \ee 
with
 \be
  U_\ell=   U+ \frac{\ell^2}{  2 b \left(A^{2}+\frac83 C^2   \right) }.
  \label{UL}
   \ee
   \end{subequations} 
   Here  $K$ is as in \eqref{A100} and $U$ as in \eqref{A6}.
    Note that the frequency  $\omega$ has disappeared from
 the system \eqref{vareq}-\eqref{Lagell},  along with $\dot \theta$.  The role of the single parameter has been taken over by the quantity $\ell$. 
We also note that equations \eqref{vareq} conserve energy:
\be
\label{H}
H=   K+U_\ell.
 \ee

\subsection{Fixed points} 

\begin{figure}[t]
 \begin{center}     
              \includegraphics*[width=\linewidth] {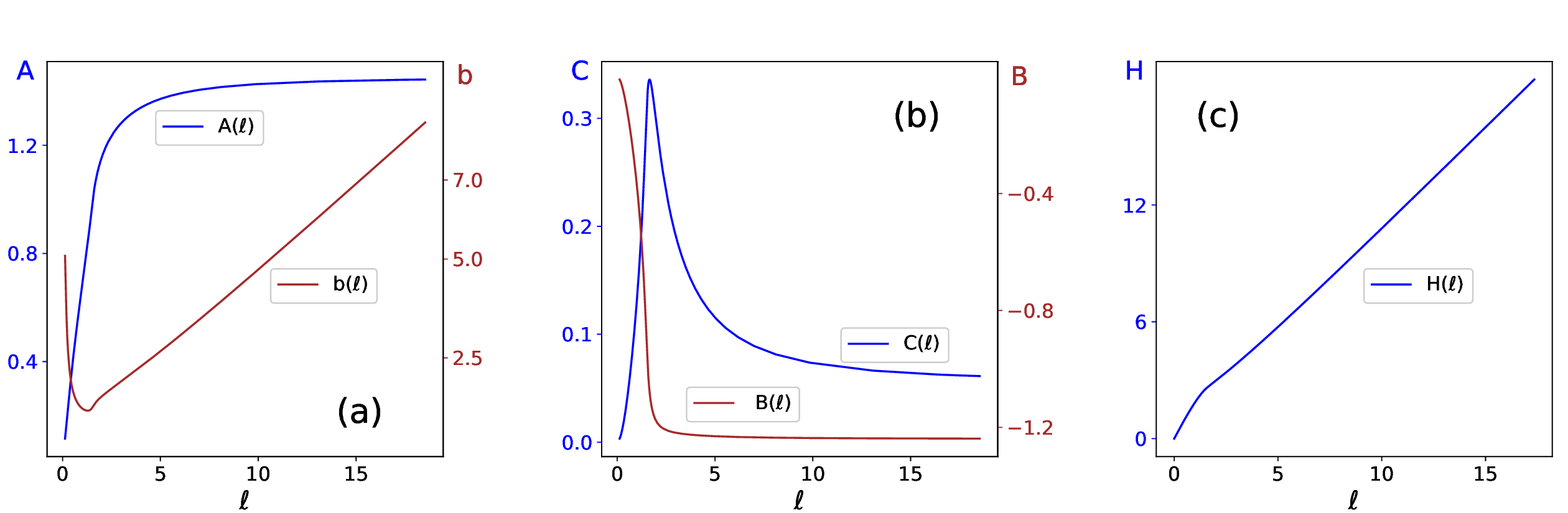}               
                                          \end{center}
  \caption{ The fixed-point solutions of the variational equations \eqref{vareq}. 
(a) amplitude of the first harmonic and width parameter; (b)  amplitudes of the zeroth and second harmonic;  (c)   energy of the fixed point.  
  \label{Fig0}  
  }
 \end{figure}

Stationary solutions of the system \eqref{vareq} are given by critical points of the function $U_\ell (A,B, C, b)$. The numerical analysis reveals that 
one such point, continuously dependent on $\ell$,
exists for all $0 \leq \ell < \infty$ (see Appendix \ref{AAA}). This nontrivial fixed point exists for the 
same reason why a classical  particle moving in a centrally-symmetric confining potential   has an equilibrium orbit
with any nonzero angular momentum. 
The components of the fixed point are presented graphically in 
Fig \ref{Fig0}(a-b). The energy of the stationary point,
\be
H(\ell)= U_\ell \left[ A(\ell), B(\ell), C(\ell), b(\ell)  \right],
\label{Hen} 
\ee
is shown in Fig \ref{Fig0}(c).

It is worth mentioning here 
 that our numerical continuation algorithm determines the components of the fixed point
as functions of $\omega$    (rather than $\ell$).  Following the branch of fixed points that extends from $\omega_0 =2$ to $\omega_c=  0.417 \, \omega_0$ 
before  reversing to $\omega_b=0.527 \, \omega_0$,
 we compute $\ell(\omega)= \omega b (A^2 +\frac83 C^2)$ (see Fig \ref{Eom} in Appendix \ref{AAA}). 
 The functions $A(\ell)$, $B(\ell)$, $C(\ell)$, $b(\ell)$ and $H(\ell)$ in Fig \ref{Fig0}
 are then plotted as parametric curves.
The asymptotic regime $\omega \to \omega_0$ corresponds to $\ell \to 0$; as the branch $\omega$ approaches its terminal point $\omega_b$, we have $\ell \to \infty$.

Returning to  the partial differential  equation \eqref{A0},  
the energy of its solutions  is given by 
\be
E [\Phi]=    \frac12   \int  \left[  \Phi_t^2+ \Phi_x^2 + (\Phi^2-1)^2   \right]  \, dx.
\label{E0}
\ee
Substituting the ansatz \eqref{A4} in the integral \eqref{E0} gives a time-dependent quantity
that cannot be used as a bifurcation measure of the trial function \eqref{A4}.
However 
when the collective coordinates $A, B, C$  and $b$  take constant values, the integral \eqref{E0} becomes periodic (with period $T$). In that case, 
 we can define 
 the {\it average\/} energy carried by  the configuration  \eqref{A4}:
 \be 
\label{E2}
{\overline E} =\frac{1}{T} \int_0^T  E[\Phi]  \, d \mathcal T_0.
\ee
Performing integration over $\mathcal T_0$ and $x$, equations
 \eqref{E0} and \eqref{E2} give
$\overline E =       U_\ell  (A,B, C, b)$,
where $U_\ell$   is as in  \eqref{UL}.
 {
Thus,
 the average field energy carried by the critical trial function (that is, the function 
 \eqref{A4} with the parameters given by the fixed point $A(\ell), B(\ell), C(\ell), b(\ell)$)
 agrees with 
 the energy integral of the variational equations \eqref{vareq} evaluated 
  at that fixed point:}
 \be
 \overline E(\ell)=H(\ell) = U_\ell \left[ A(\ell), B(\ell), C(\ell), b(\ell) \right].
 \label{E3}
 \ee

\subsection{Stability of the fixed point}
\label{stability} 

In order to classify the stability of the fixed points, it is sufficient to examine 
 perturbations preserving the integral $\ell$.
 Linearising
 equations \eqref{vareq} 
 and
 assuming the time dependence of the form 
\be
( \delta A, \delta B, \delta C, \delta b)^T= e^{( \lambda / \epsilon) \mathcal T_1} {\vec y},
\label{pert}
\ee
where ${\vec y}$ is a constant 4-vector,
gives a generalised eigenvalue problem
\be
M {\vec y} = - \lambda^2 P  {\vec y}.
\label{MP}
\ee
Here $M$ and $P$ are $4 \times 4$ symmetric real matrices.
The $M$ matrix is given in the Appendix \ref{BBB} while 
\be
P = \left( 
\begin{array}{cccc} 
2b & 0 & 0 & A \\
0 & \frac83 b & 0 & \frac43 B \\
0 & 0 & \frac43 b  & \frac23 C \\
A & \frac43 B & \frac23 C & P_{44}
\end{array}
\right),
\label{P}
\ee
\[
P_{44}= \pi^2 \frac{5A^2 + 16B^2 + 8C^2}{90b} + \frac23 \frac{A^2}{b}.
\]
Note that the ansatz \eqref{pert} tacitly assumes the exponents $\lambda$ being of order $\epsilon$.

The determinant of $P$ is given by
\be
\label{detP}
\det P= \frac{32 b^2}{405}   \left[ 5(\pi^2+3) A^2+ 2 (4 \pi^2-15) (2B^2+C^2) \right].
\ee
Since $\det P>0$,  Sylvester's criterion implies that  $P$ is positive definite. Therefore, all generalised eigenvalues $ \lambda^2$ are  real
and the exponents  form pairs of real or pure imaginary opposite values, $\lambda$ and $-\lambda$.

\begin{figure}[t]
 \begin{center} 
               \includegraphics*[width=0.7\linewidth,height=80mm] {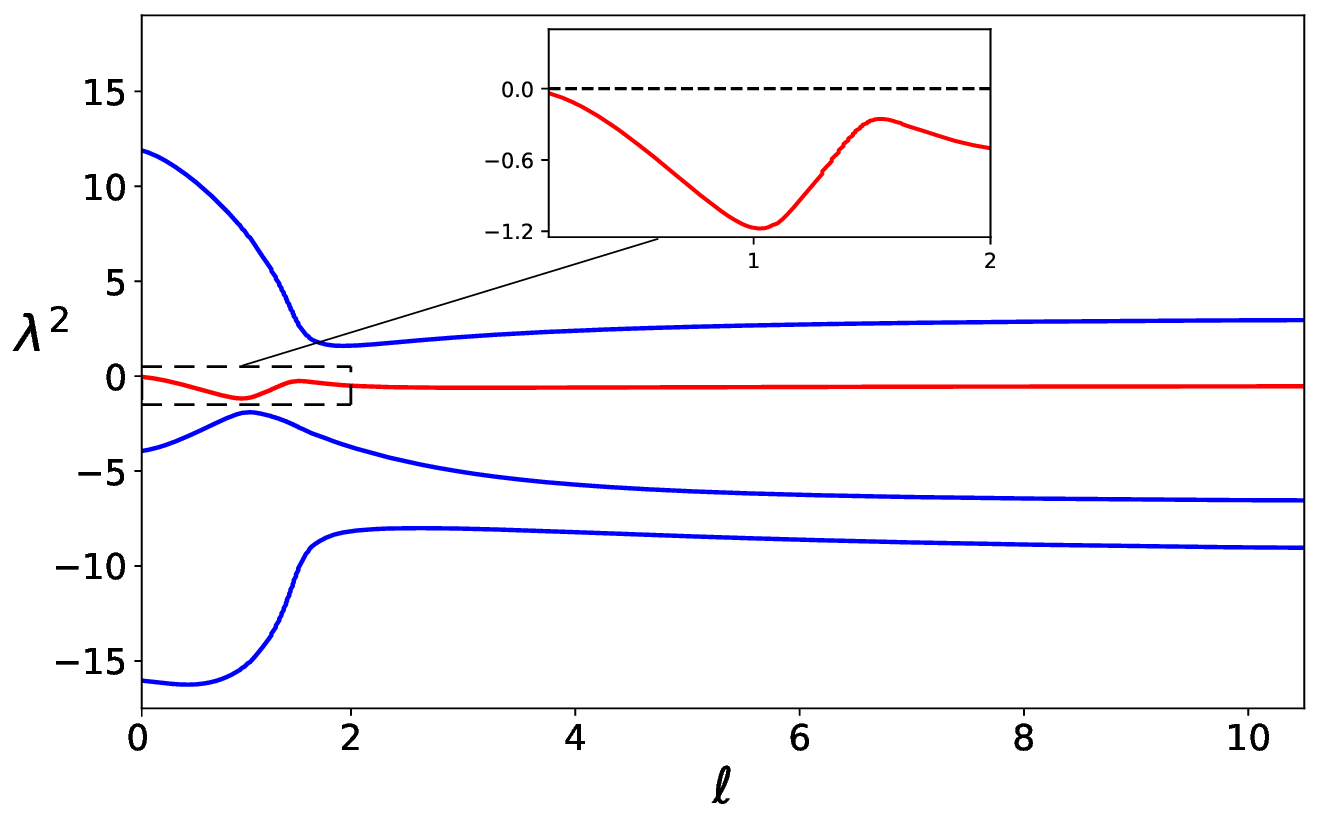}                                          
                \end{center}
  \caption{\label{EVs} 
  Roots of the characteristic equation $\det (M  + \lambda^2  P)=0$.
The generalised eigenvalues $\lambda^2$ are shown as functions of $\ell$. The blow-up of a segment of the branch emerging from the origin (plotted in red)
highlights the fact that the eigenvalue never returns to zero.
  }
 \end{figure}

The generalised eigenvalues are computed numerically; the results are in Fig \ref{EVs}. 

 {
A negative eigenvalue emerges from the origin as $\ell$ is increased from zero.  
This is the only $O(\epsilon^2)$ eigenvalue present;
all other generalised eigenvalues shown in  Fig \ref{EVs} (including three other ones occurring for small $\ell$) 
are of order 1. Those large eigenvalues are inconsistent with the assumption underlying the ansatz \eqref{pert};
they do not carry any information on the stability properties of the oscillon.
By contrast, the negative $O(\epsilon^2)$ eigenvalue appearing in the small-$\ell$ regime 
agrees with 
the stability of the  Dashen-Hasslacher-Neveu's periodic solution  in the limit $\omega \to \omega_0$
(where its slowly-varying amplitude satisfies the nonlinear Schr\"odinger equation.)  
}

 {
The absence of $O(\epsilon)$ exponents outside the asymptotic regime $\omega \to \omega_0$ implies that the 
oscillon remains stable  as $\omega$ is reduced to lower values ---
 for as long as our variational approximation remains valid. Indeed, had the instability set in at some $\omega_\mathrm{cr}< \omega_0$, 
 it would have brought along slowly-varying amplitude perturbations (of the periodic oscillation with the frequency $\omega_\mathrm{cr}$). 
 The associated $O(\epsilon^2)$ eigenvalues would have been captured by the eigenvalue problem \eqref{MP}.
}

\section{Numerical standing waves}
\label{numerical}
\subsection{Energy-frequency diagram} 

To assess the accuracy of the variational approximation \eqref{A4},  we consider standing-wave solutions 
of  the equation \eqref{A0}. The standing waves (also known as nanopterons \cite{Boyd2} and quasibreathers \cite{Fodor2})
are temporally periodic solutions assuming prescribed values at
 the ends of the  finite interval 
$-R \leq x \leq R$. 
Confining the analysis to spatially symmetric (even) standing waves, we determine these as solutions of 
a boundary-value problem posed 
on a rectangular domain $(0,R) \times (0,T)$.
Equation \eqref{A0} with the 
boundary conditions
\be 
\label{BC}
\Phi_x(0,t) =0; \quad   \Phi(R,t)=1; \quad \Phi(x,0)= \Phi(x,T)
\ee
 was solved by a path-following algorithm with a Newtonian iteration. 
 The half-length of the     interval,  $R$, was set to $40$. 

A typical solution consists of a localised core and a non-decaying small-amplitude wing, resulting from the interference of the 
outgoing radiation and radiation reflected by the boundary at $x=R$.

\begin{figure}[t]
 \begin{center} 
            \includegraphics*[width=0.9\linewidth] {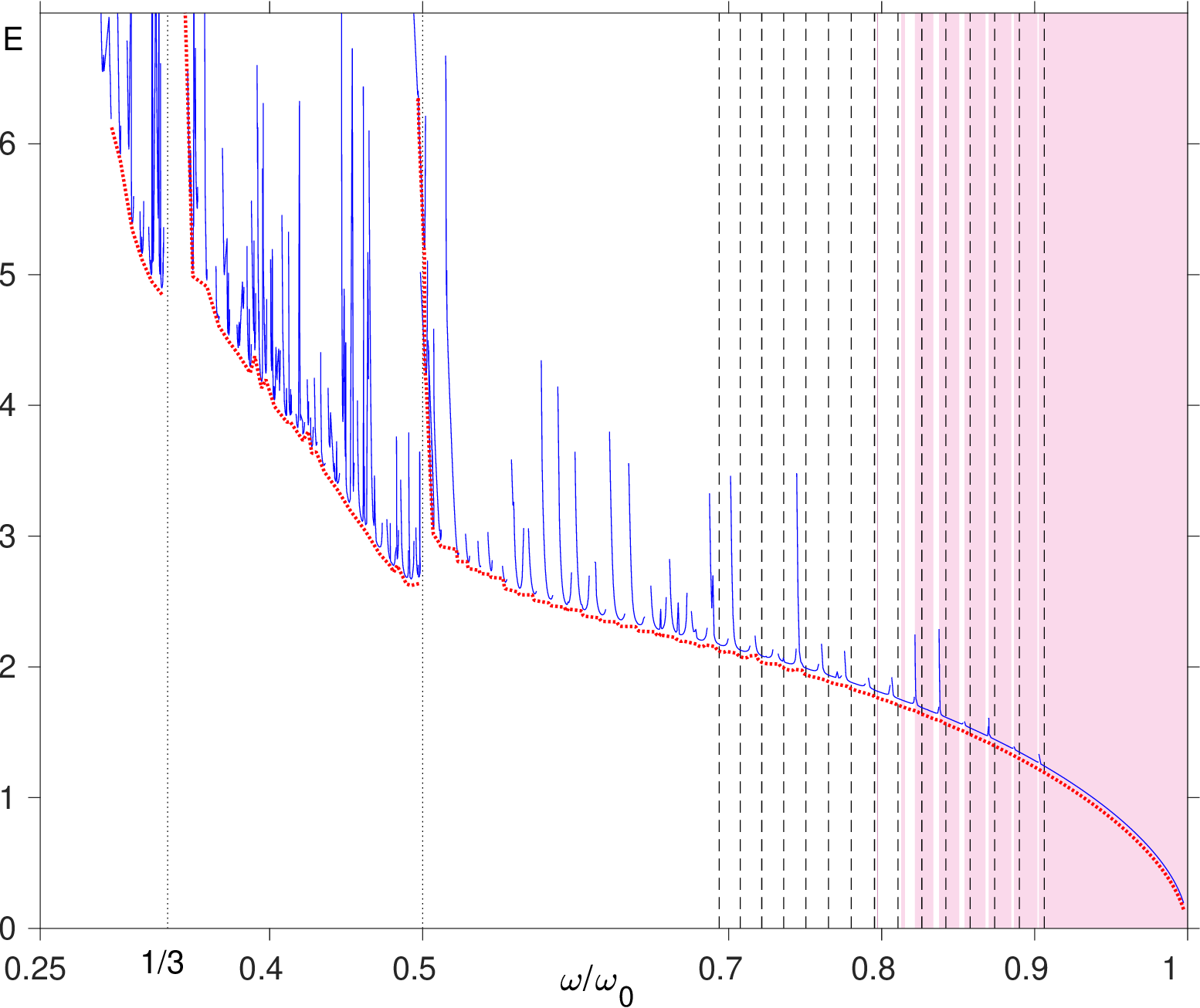}   
                     \end{center}
  \caption{   \label{Fig3}  
   The energy \eqref{ER} 
    of  the numerical standing-wave solution 
     (blue).
   The vertical dashed lines are drawn at the subharmonic resonance frequencies $\omega=\Omega^{(n)}/2$,
   where $\Omega^{(n)}$  are the frequencies \eqref{G16} 
   of 
   the linear waves.
 The red dotted arc underlying the numerical
  curve is the envelope of the family of spikes. (For visual clarity, it has been shifted down by a tiny amount from its actual position.)
  Clearly visible are discontinuities of the envelope at $\omega/ \omega_0= 1/2$ and $\omega/\omega_0=1/3$. The pink-tinted bands demarcate the frequency intervals 
  where all Floquet multipliers lie on the unit circle.     }
 \end{figure}

Fig \ref{Fig3} shows the energy 
\be
E_R [\Phi]=       \int _0^R \left[  \Phi_t^2+ \Phi_x^2 + (\Phi^2-1)^2   \right]  \, dx      
\label{ER}
\ee
of solutions of the boundary-value problem \eqref{A0}, \eqref{BC}. There are two features that attract attention in the figure.

First, 
the energy-frequency diagram exhibits what appears to be a sequence of spikes. 
On a closer examination, each     \enquote{spike}   turns out to consist of a pair of $E(\omega)$ branches rising steeply but not joining together. 
As the point $(\omega, E(\omega))$ climbs up the slope of a spike, the amplitude of the standing wave's wing grows --- this accounts for the rapid growth of 
the energy of the solution.

 {Each spike is produced by the resonance between an overtone of the
frequency of the core of the standing wave
and the eigenfrequency $\Omega^{(n)}$ of a  linear mode with some $n$.   The linear modes  are given by}  
 \be
 \label{G2} 
 \Phi= 1+    \epsilon  \sin (\Omega^{(n)}  \mathcal T_0)   \cos (\kappa^{(n)} x ),
 \ee
 where 
     \be
 \Omega^{(n)}= \sqrt{\omega_0^2+ \left( \kappa^{(n)}\right)^2},     \quad     \kappa^{(n)}= \frac{\pi}{R} \left( n+ \frac12 \right)  \quad (n=0,1,2, ...).
 \label{G16} 
 \ee
(Similar resonances have been detected in the three-dimensional version of the $\Phi^4$ model
\cite{Alex_PRD}.)
The positions of the $\frac12$  undertones of the linear modes are marked by the vertical dashed lines in Fig \ref{Fig3}. 
One can clearly see a correspondence between the positions of the   \enquote{spikes}   and the points $\omega=\Omega^{(n)}/2$ through which the vertical lines are drawn. 
 {
(We only show the mode undertones in the right-hand portion of the diagram, where the spikes are thin and stay clear of each other.)
}

 The positions of the spikes are sensitive to the choice of the interval half-length, $R$. Let $\omega$ be a fixed frequency  with $\omega_0/2 < \omega <\omega_0$
 and $R_A$  an arbitrarily chosen half-length ($R_A \gg 1$). Denote $k$ the wavenumber of the second-harmonic radiation, satisfying the dispersion relation 
 \be
\label{nth}
(n \omega)^2 = \omega_0^2+  k^2
\ee 
 with $n=2$. 
 By tuning $R$ to a suitably chosen value $R_\omega$ within the interval  $(R_A, R_A + \pi/ k)$, 
  the amplitude of the      \enquote{wing}   can be minimized (yet not reduced to zero).  The corresponding value  of $E_R$ gives the minimum
  energy of the family of standing waves with frequency $\omega$: $E_{\rm min}(\omega)=  E_{R_\omega}$.
  The graph of $E_{\rm min}(\omega)$  comprises segments of the $E_{R_A}(\omega)$ curve outside the neighbourhoods of the spikes, while
 the full, gapless,   $E_{\rm min}(\omega)$ arc can be obtained as the envelope of the family of  curves with $R$ in $(R_A, R_A + \pi/ k)$.

Another aspect of   Fig \ref{Fig3} that is worth  commenting on, concerns the fragmentation 
of the $E(\omega)$ curve into three disjoint branches with $\omega$ in
 the intervals $(\omega_*, \frac13 \omega_0)$,      $( \frac13 \omega_0,   \frac12 \omega_0)$ 
and $( \frac12 \omega_0, \omega_0)$, respectively.
Here  $\omega_* =0.289 \,  \omega_0$
is the lowest  frequency value  for which we obtained a solution of the boundary-value problem \eqref{A0}, \eqref{BC}.
The endpoints of the intervals, $\omega_0/2$ and $\omega_0/3$, are marked by a rapid growth of the energy $E_{\rm min}$
as $\omega$ approaches these points from the right,
and by finite energy jumps --- as $\omega$ approaches these values from the left.

 {The fragmentation admits a simple explanation in terms of the dispersion relation \eqref{nth}.}
According to \eqref{nth}, the $n$-th harmonic radiation can only be emitted by the core oscillating with the frequency $\omega>\omega_0/n$. 
In the event of the coexistence of the $n$-th and $(n+1)$-th harmonics, the lower (the $n$-th) overtone will be energetically dominant. 
Therefore the radiation waves should be dominated by the fourth, third and 
second harmonic 
 in the  frequency ranges   $(\frac{1}{4} \omega_0,   \frac13 \omega_0)$,
  $(\frac13 \omega_0,   \frac12 \omega_0)$ and   $(\frac12 \omega_0, \omega_0)$, 
 respectively.
 The  harmonic waves forming the wings of our numerically obtained nanopterons do comply with  this selection rule.

 Raising $\omega$ through $\omega_0/2$ opens a channel of the quadratic radiation --- a more powerful channel than 
 the $n=3$ channel available to cores oscillating at  $\omega< \omega_0/2$.
 The amplitude of the wing increases and this accounts for the energy jump in Fig \ref{Fig3}. Similarly, 
 the energy jump observed as $\omega$ is increased through $\omega_0/3$ is due to the opening  %  turning on 
 of the third-harmonic channel, 
 not accessible to the standing waves with $\omega < \omega_0/3$. We note that 
 the energy jumps have the same origin as the staccato flashes of radiation from a slowly fading oscillon in the deformed
 sine-Gordon  and $\phi^6$ equations \cite{Dorey,Nagy}.

  {
As  $\omega$ approaches $ \omega_0/2$  from above, the wavenumber of 
 the second-harmonic radiation tends to zero.  The radiation wave
transforms into a spatially unifom oscillation
which acts  as a parametric driver on the core of the standing wave.
Since the frequency of the driver is double that of the core, the parametric resonance condition is met and the amplitude of the
core is boosted \cite{BBK}.
  A similar argument explains the energy growth occurring as  $\omega$ approaches   $\omega_0/3$ from above.
}

\subsection{Stability of standing waves}

 {
Let $\Phi_\omega(x,t)$ denote the solution of the boundary-value problem \eqref{A0}, \eqref{BC} with $\omega = 2 \pi/T$.
To classify its stability, we linearise \eqref{A0} about this solution:}
\be
y_{tt}-y_{xx} -2y  + 6 { \Phi^2_\omega}(x,t)  y=0.
\label{S1}
\ee
Equation \eqref{S1} is supplemented with the boundary conditions
\be 
y_x(0,t)= y(R,t)=0; 
\label{C2}
\ee
that is, we confine our study to   perturbations sharing the symmetry of the standing wave and 
vanishing at the same point on the $x$-line. 

Having expanded $y(x,t)$ in the cosine Fourier series  in the interval $(0,R)$ and keeping only the first $N$ harmonics,
we have evaluated the monodromy matrix of the $T$-periodic solution for each $\omega$ in Fig \ref{Fig3}.
(We took $N=512$.) 
  If all eigenvalues $\mu_n$  ($n=1,2, ..., 2N$)
  of the monodromy matrix satisfy
 $|\mu_n|=1$,   the periodic solution $\Phi_\omega(x,t)$ is deemed stable.
 If there are Floquet multipliers with $|\mu_n|>1$, the standing wave $\Phi_\omega$  is  linearly unstable.
 % however the evolution of the unstable  perturbations may give rise to  stable localised solutions that are close to $\Phi$ in its core region.
% however, the growth of the unstable perturbations should not necessarily 
% produce a noticeable deformation of the wave's core.

The frequency intervals with no multipliers outside the unit circle are indicated by pink bands in Fig \ref{Fig3}. In particular, 
the entire region $0.903 <  \frac{\omega}{\omega_0} < 1$ is found to be stable. 
 As $\omega$ is continuously turned down from $0.903 \, \omega_0$,
a pair of real eigenvalues ($\mu_m$ and $\mu_m^{-1}$)  repeatedly
 emerge and return to the unit circle.
(The intervals of $\omega$ characterised by the presence of $\mu_m >1$ 
 are left blank in Fig \ref{Fig3}.) After the frequency has reached below $0.798 \, \omega_0$, 
the off-circle pair remains in the Floquet spectrum for any further decrease of $\omega$.

The instability occurring in parts of the  range $\omega \gtrsim 0.75 \, \omega_0$ is weak. 
(Here we are assuming that the solution is not on the slope of a resonant peak).
 Specifically,  the  growth rate $\lambda= \frac{1}{T} \ln \mu_m$ associated with the Floquet multiplier $\mu_m>1$
 is bounded by $3 \times 10^{-3}$. 
As $\omega$ is decreased below $0.75 \, \omega_0$, the unstable real eigenvalue becomes larger. In addition, complex quadruplets $\{ \mu, \mu^*, \frac{1}{\mu}, \frac{1}{\mu^*}  \}$ 
emerge from the unit circle.

To understand the effect of instability, we solved equation \eqref{A0} on the interval $(-R, R)$,
with the initial conditions
  {
\be
\Phi(x,0) = \Phi_\omega(x,0),
\quad
\frac{d \Phi (x,0)}{dt} = \frac{d \Phi_\omega (x,0)}{dt}, \quad |x| \leq R
\label{IC1}
\ee
and  boundary conditions $\Phi(\pm R,t)=1$.
%  In all cases that we examined, the evolution of the instability affected the amplitude and phase of the
%   wing of the nanopteron,
% but it never led to any significant deformation of its core. 
% Since the resulting changes would not be noticeable in most physical settings, we are referring to the standing-wave solutions of  equation \eqref{A0}   % as {\it practically stable.}
The initial perturbation of the standing wave was  introduced by the finite-difference approximation 
of the derivative $d \Phi_\omega(x,0) /dt$ of the boundary-value problem \eqref{A0}, \eqref{BC}.
}

 {We examined the standing waves with a string of $\omega$ values from the interval $(0.45 \,  \omega_0,$
 $0.90 \, \omega_0)$. 
The growth of instability  of an unstable nanopteron with $0.75 \, \omega_0 < \omega  <0.90  \, \omega_0$ was expectedly very slow; 
the structure had to perform hundreds or even thousands of oscillations before any change would become noticeable.
Consequently, 
 we classify  the standing-wave solutions  in that interval  as {\it practically stable} and turn to the 
 complementary range. }

  {
When a nanopteron with the frequency in $(0.45 \,  \omega_0, 0.75 \, \omega_0)$
 had its energy close enough to the envelope value $E_{\mathrm{min}}(\omega)$
(that is, when we picked up a standing wave with the energy lying at the bottom of one of the lobes in Fig \ref{Fig3}), 
the growing instability would not lead to the desintegration of the localised structure.
 It would just send the standing wave to the stable part of the $E_{\mathrm{min}}(\omega)$ curve
(typically, to $\omega$ values near $ 0.85 \, \omega_0$).
}

          \begin{figure}[t]
 \begin{center} 
                                     \includegraphics*[width= \linewidth] {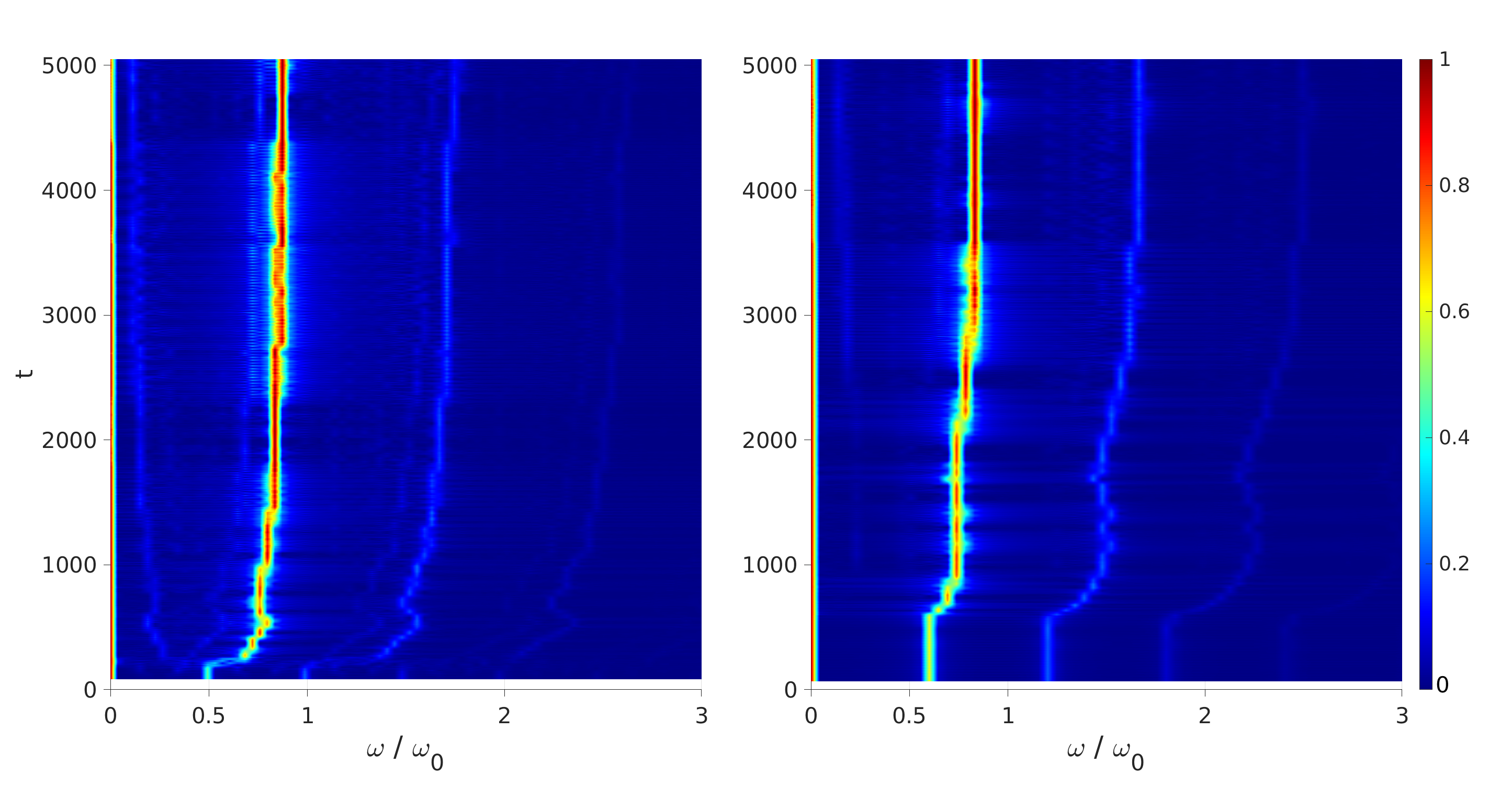}       %                                   \includegraphics*[width= \linewidth] {fig4_3.eps}    
                              \end{center}
  \caption{   \label{FigNew}    {
  Spectral evolution of the core of the unstable standing wave.  Colour encodes 
  the (normalised) absolute value  of the short-time Fourier transform $F(\omega,t)=\int_t^{t+\tau} e^{i \omega t^\prime} \Phi(0,t^\prime) dt^\prime$,
  where $\Phi(x,t)$ is the 
  solution of the initial-value problem  \eqref{A0}, \eqref{IC1}
   with $\omega=0.4975 \, \omega_0 $ (a) and $\omega = 0.606 \, \omega_0$ (b). 
In either case a rapid growth of the fundamental frequency of the oscillatory structure
gives way to  its slow drift along the $E_{\mathrm{min}}(\omega)$ curve. 
The width of the sliding window, $\tau$,  was set to 83 in (a) and 69 in (b).
}
}
 \end{figure}

 {
 The transition would proceed in two stages. (See Fig \ref{FigNew}.)
The first stage involves a rapid growth of the fundamental frequency  of the oscillatory structure, with its energy dropping under the envelope curve in
Fig \ref{Fig3}. The second part of its journey begins when the growing frequency brings
 the point $(\omega, E(\omega))$  back to the envelope curve (whose descent becomes steeper as $\omega$ grows). 
 Once the structure has caught up with the $E_{\mathrm{min}}(\omega)$ curve,
   it will continue to lose energy --- slowly
sliding down that arc. 
}

 {
It is appropriate to note that there is no fundamental difference between the evolution of the unstable standing waves 
with $\omega$ on the left and on the right of the discontinuity at
$\omega_0/2$ in Fig \ref{Fig3}
(as long as $\omega$ does not fall under $0.45 \, \omega_0$).  (Compare Fig \ref{FigNew}(a) to  \ref{FigNew}(b).) Furthermore, the  instability growth is not accelerated as $\omega$ crosses through the critical value
of $\omega_0/2$. Unlike the staccato transition reported in  \cite{Dorey,Nagy}, the instability does not depend on the 
availability of radiation channels.
}

 {
We close this section with a remark on the standing waves with low frequencies, $\omega < 0.45 \,  \omega_0$.  
The instability takes violent forms in this range:  a tiny perturbation breaks the standing wave 
into a kink-antikink pair or a pair of diverging oscillons. The analysis of these processes is left as a future challenge. 
}

\subsection{Standing waves vs oscillons}

 {
What do the standing waves in a box teach us 
 about the behaviour of  oscillons, 
the oscillatory radiating lumps on the endless line? To gain insight into the relation between the
two types of objects, we simulated equation \eqref{A0} with the initial condition of the form
\begin{align}
\Phi (x,0)= \Phi_\omega(x,0), \quad 
\frac{d  \Phi (x,0)}{dt} = \frac{ d \Phi_\omega(x,0)}{dt}, \quad  |x|  \leq R; \\
\Phi (x,0)= 
\frac{d \Phi (x,0)}{dt} =0 , \quad  R< |x| < \tilde R,
\end{align}
on an interval $(-\tilde R, \tilde R)$, where  $\tilde R \gg R$. (We took $\tilde R=80$.)
 The radiation-absorbing pads placed near the endpoints of the interval  emulated the infinite line situation. 
}

 {In this ``infinite space" experiment, 
a standing wave that had been classified as stable in the $-R <x <R$ box was observed to   drift, adiabatically, 
down
 the $E_{\mathrm{min}}(\omega)$ curve in Fig \ref{Fig3}.
 By contrast,  an {\it unstable\/} nanopteron was seen to leave the curve for a short while, 
  return  to it at a higher frequency, and 
 then start sliding down the arc --- first due to the growth of instability and later, when it is already in the stable domain,
 due to the radiative energy losses. 
 }

 {
Thus, the family of standing waves with a varied  frequency in a long box provides the trajectory 
followed by an adiabatically fading oscillon in its phase space.
}

\section{Numerical solution vs  variational approximation}
\label{comparison} 

\begin{figure}[t]
 \begin{center}  
   \includegraphics*[width=0.6\linewidth] {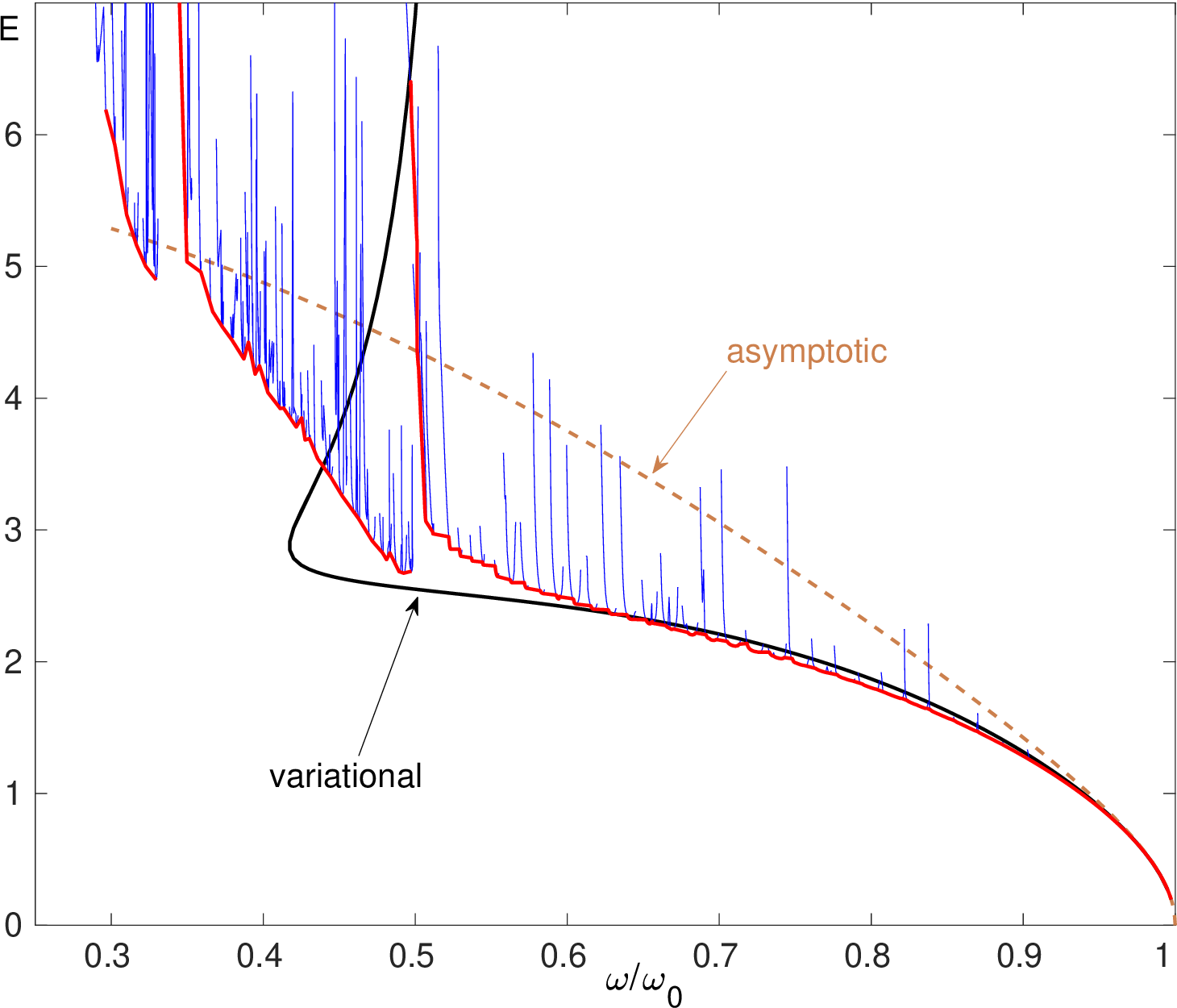}     
                     \end{center}
  \caption{   \label{Fig4}
   The energy \eqref{E0} 
    of  the standing-wave solution determined numerically 
     (blue), with the envelope curve plotted in red.
     (Unlike Fig \ref{Fig3}, the envelope is shown in its true position here.) 
     The black line depicts the average energy \eqref{E3}  carried by 
   the trial function \eqref{A4} with parameters 
   given by the fixed-point  of the variational equations.
   The dashed parabola indicates the asymptotic result \eqref{E1}. 
  }
 \end{figure}

Fig \ref{Fig4} compares the energy of the standing wave 
  with the average energy \eqref{E3} carried by the critical trial function. 
  (The latter is given by \eqref{A4} where $A,B, C$ and $b$ are taken to be the components of the fixed-point solution 
 of  equations \eqref{vareq}.) 
 For further comparison, we have included the energy of the asymptotic solution \eqref{A1},
\be
E=\frac43 \epsilon + \frac{32}{81} \epsilon^3,
\label{E1}
\ee
where $\epsilon = \sqrt{\omega_0^2-\omega^2}$ and  corrections higher than $\epsilon^3$ have been disregarded.

In the frequency range $(0.6 \,  \omega_0, \omega_0)$, the variational  result \eqref{E3} provides a fairly accurate approximation of the 
envelope of the numerical $E(\omega)$ curve.  Remarkably, the variational answer is much closer to $E_{\mathrm{min}}(\omega)$ 
than the asymptotic 
expansion \eqref{E1}.

 As $\omega$  is reduced from $0.6 \, \omega_0$ to $0.5 \, \omega_0$, the variational  $\overline E(\omega)$ dependence deviates from its numerical counterpart
and once  $\omega$ has fallen under $\omega_0/2$, the numerical-variational correspondence breaks down entirely. 
The envelope of the numerical curve in Figs \ref{Fig3} and \ref{Fig4} is split into three fragments, with $E_{\mathrm{min}}$ growing monotonically 
as $\omega$ decreases within each fragment. There is a unique value of $E_{\mathrm{min}}$ for each $\omega$. 
By contrast, the variational  energy  $\overline E(\omega)$ grows as $\omega$ is reduced from $\omega_0$  to $\omega_c=0.417 \, \omega_0$,
but then the curve turns back so that there are  two coexisting branches for each $\omega$ in the interval $(\omega_c, \omega_b)$,
where $\omega_b=0.527 \omega_0$.

The stability of the standing wave with $\omega >0.9 \, \omega_0 $ and its {\it practical\/}  stability
in the interval  $0.75 \, \omega_0 < \omega< 0.9 \, \omega_0$ 
are  reproduced by the 
variational approach.  
 {However the stability properties are not immune from
the general deterioration of the variational approximation occurring as $\omega$ is reduced.
Specifically, the numerical analysis reveals that the nanopteron becomes unstable as $\omega$ falls under $0.75  \,\omega_0$
 whereas 
the variational method fails to capture this instability. 
}

The numerical analysis suggests three factors that 
 contribute to the failure of the variational approximation with low $\omega$.
First, the amplitude of the radiation emitted from the core of the wave (and reflected from the boundary at $x=R$) 
grows as its frequency is decreased.   By contrast,  the ansatz \eqref{A4} does not take into account the radiation wing at all.
Second, as $\omega$ is lowered below $\omega_0/2$,  the contribution of the first harmonic
to the Fourier spectrum of the core decreases while the role of higher harmonics grows. 
% As the frequency falls under $0.45 \, \omega_0$, 
%  the amplitude of the second harmonic becomes larger than, and the amplitude of the third harmonic comparable to,  the amplitude of 
% the first harmonic. When $\omega$ is under $0.4 \, \omega_0$, both second and third harmonics make larger contributions to the dynamics than the first harmonic.
As a result, the variational ansatz \eqref{A4} --- which  keeps $\cos ( \omega \mathcal T_0+\theta)$ but
does not include $\cos 3 (\omega \mathcal T_0+\theta)$ --- becomes inadequate. 
 {Finally, the spatial profile of the low-frequency nanopteron's core exhibits several humps 
and cannot be emulated by the bell-shaped ansatz \eqref{A4}.
}

\section{ Concluding remarks} 
\label{conclusions}

A variational approach to a localised structure aims to identify the collective coordinates that capture
the essentials of its dynamics. 
% The agreement between the finite-dimensional description of the structure  and its true properties revealed by the numerical analysis validates the set of collective variables. 
The choice of essential coordinates is validated by the agreement between the finite-dimensional description of the structure
and its true properties as revealed by the numerical analysis.
% By contrast,   the lack of agreement disqualifies the choice of the trial functions. 
In this study, we utilised the  multiscale variational method to determine the nonlinear modes 
responsible for the formation and stability 
 of the one-dimensional $\phi^4$ oscillons.

  The method was previously applied  \cite{BA}
  to  oscillons of the Kosevich-Kovalev  model \cite{KK},
   defined by the Lagrangian 
 \be 
L= \frac12 \int  \left(   \phi_t^2-\phi_x^2 - 4 \phi^2 + \phi^4 \right) dx.
\label{KK}
\ee
An ansatz comprising three variables,
 \be
 \phi= A \cos (\omega \mathcal T_0 + \theta) \,  \mathrm{sech} \left (\frac{x}{b} \right),
 \label{J2}
 \ee
 was found to provide an accurate agreement with the numerical simulations of the corresponding equation of motion.
 
The $\phi^4$ Lagrangian of the present paper,
 \be 
L= \frac12 \int  \left(   \phi_t^2-\phi_x^2 - 4 \phi^2 -4 \phi^3 -   \phi^4 \right) dx,
\label{J4}
\ee
is different from \eqref{KK} in the presence of the symmetry-breaking $\phi^3$ term.  This term would not contribute to the
effective Lagrangian generated by the trial function \eqref{J2};  the single-harmonic ansatz   is    \enquote{blind} to  the cubic term.
To capture the asymmetry in oscillations,  we have had to extend the single-harmonic ansatz \eqref{J2} to a sum of three harmonics in \eqref{A4}.  
Accordingly, the number of collective variables  has increased from three to five.

Our variational study was complemented by the numerical analysis  of  equation \eqref{A0}  on a periodic two-dimensional  domain 
$(0,R) \times (0,T)$.  
Solutions of this boundary-value problem are standing waves consisting of a localised core and small-amplitude wings 
formed by the interference of the radiation emitted by the core and radiation reflected from the boundaries.
 % To assess the accuracy of the variational approximation, we have solved equation \eqref{A0} numerically, on a periodic two-dimensional  domain  $(0,R) \times (0,T)$. 
 We obtained solutions with all $\omega$  between $0.3  \, \omega_0$ and $\omega_0$, where $\omega = 2\pi/T$ is the fundamental frequency of the wave 
 and $\omega_0$ is the endpoint of the continuous spectrum.

Results are summarised in Fig \ref{Fig3} which shows the energy of the standing wave as a function of its frequency.
%The standing wave  consists of a localised core and small-amplitude wings formed by the interference of the radiation emitted by the core and radiation reflected from the boundaries.
The diagram features a sequence of spikes generated by the resonant growth of the wing amplitude. 
The envelope of the family of $E_R(\omega)$ curves with varied $R$ gives the energy curve of the standing wave with the thinnest wings.
The envelope is found to be fragmented according to the 
dominant radiation harmonic (second, third, or higher).
%Another noteworthy conclusion of the numerical analysis is  the stability or practical stability of standing waves 

Fig \ref{Fig4} attests to a good agreement between the numerical energy-frequency dependence
and its variational counterpart $\overline E(\omega)$ in the interval $(0.6 \, \omega_0,     \omega_0)$. 
%There also is consistency between the stability of the  variational fixed point 
% and stability (or practical stability) of  the 
%  standing waves with frequencies  between
% $0.75 \, \omega_0$ and $\omega_0$.
As $\omega$   is decreased below $0.6 \, \omega_0$, the 
agreement deteriorates and then breaks down completely. 
While the energy of the numerical solution changes monotonically 
within each of the three fragments of its $E(\omega)$ diagram, the variational curve turns back into a 
coexisting branch of fixed points (Fig \ref{Fig4}). 
% It remains a challenge to determine whether the standing-wave counterpart of the  coexisting branch actually   exists.

 {
As for the stability properties of the standing wave and its finite-dimensional approximation, their interval of consistency is even shorter.
The numerical standing wave loses its stability once $\omega$ has dropped below $0.75 \, \omega_0$ 
whereas the  fixed-point solution of the variational equations remains stable over its entire existence domain, $(0.417 \, \omega_0, \omega_0)$. 
}

\acknowledgments
We thank N Quintero for useful discussions.
This research was supported by the NRF of South Africa (grant
 No  SRUG2204285129). 
 % and  a  bilateral collaborative grant from the NRF and   Joint Institute for Nuclear Research (Grant UID 120467). 
One of the authors (A.D.) gratefully acknowledges 
 a visiting fellowship from the Joint Institute for Nuclear Research  where this project was completed.

\appendix
\section{Appendix: Fixed points of the variational equations}
\label{AAA}

The fixed points of the dynamical system \eqref{vareq} 
satisfy  four simultaneous algebraic equations
\begin{subequations} \label{fp} 
\begin{align}
\frac{1}{3b^2}-1 + A^2+ \frac45 C^2 + \frac45 \left( \frac52 +2B+C  \right)^2=\omega^2,   
	\label{A13} \\
	\frac13 \left( 5+\frac{1}{b^2} \right) B+ \frac54 A^2 + 2B^2+   C^2+  \frac12 A^2(2B+C) +\frac27 B (2B^2+3C^2)=0,
	\label{A14}    \\
		\left(5+  \frac{1}{b^2} \right)C + \frac{15}{4}  A^2 + 12    BC + 3 A^2(B+C) + \frac{9}{7} (4B^2+C^2)C= 5\omega^2 C, 
		\label{A15}     \\
	\frac58 \left( 4- \frac{1}{3b^2} \right) A^2  +\frac13 \left( 5-\frac{1}{b^2} \right) (2B^2+C^2)
	+ \frac52 A^2( 2B+C) + \frac43B (2B^2+ 3C^2)    \nonumber  \\
	+ \frac{5}{16} A^4 + \frac{1}{14} (8B^4 +3C^4 +24 B^2C^2) + A^2 (2B^2+C^2+2BC)= \frac{5}{24}  \omega^2 \left( 3 A^2+ 8 C^2 \right). 
			\label{A16} 
		\end{align}    \label{FPs}
	\end{subequations}
	%In \eqref{fp}, 
	Note that we have switched from the 	 parametrisation 
	by  $\ell$ back to the frequency $\omega$, where
	\be
\omega= \frac{\ell}{b \left( A^2 + \frac83 C^2 \right)}.
\label{ome}
\ee
	While the invariant  $\ell$  characterises  the dynamics (in particular, stability) 
of solutions to equations \eqref{vareq},   $\omega$
appears to be a more convenient parameter in the 
 search for roots of \eqref{FPs}.

Expressing $A^2$ from equation \eqref{A13} 
and substituting it in  \eqref{A14},  we obtain
 \be
  b^{-2}=     \frac { f_\omega(B,C) }{ C+ \frac52    }
 \label{A24}
 \ee 
 with
 \begin{align}
 f_\omega = 10B+ 12B^2+6C^2+ \frac{12}{7} B(2B^2+3C^2)    \nonumber \\ +
 3 \left( \frac52        +2B+C  \right) 
 \left[ \omega^2+1-\frac45 C^2 -\frac45 \left(\frac52  +2B+C \right)^2    \right].  \label{f}
\end{align} 
A similar substitution of $A^2$ in \eqref{A15} yields
\be
 b^{-2}=\frac{ g_\omega(B,C)}{B+  \frac54       }
\label{A25}
\ee
with
\begin{align}
g_\omega= \frac32(1+ \omega^2) \left(\frac{5}{2} + 2B- \frac43 C \right)  +   \left[  10+12B + \frac97(4B^2+C^2 )  \right]C \nonumber   \\
-\frac{6}{5} \left( \frac52 +2B+2C  \right) \left[ \left(\frac52+ 2B+C \right)^2 +C^2 \right],
\label{g} 
\end{align}
%We note that $b^{-2} \to \infty$ (and therefore, $A^2 \to \infty$) as  $E \to -\frac{10}{4}$ and as $B \to -\frac54$.
while eliminating $A^2$ between \eqref{A13} and 
 \eqref{A16} produces
\be
b^{-4} + p_\omega b^{-2}   +q_\omega  =0,
\label{A18}  
\ee
where
\be
p_\omega(B,C)= 2 \left[ \frac{16}{5} \left( B+ \frac54 \right) \left( C+ \frac52 \right) - \omega^2 -6  \right]  
\label{p} 
\ee
and 
\begin{align} 
q_\omega(B,C)= 
16 \left[ 2B^2+(1-\omega^2)C^2 +\frac45 B(2B^2+3C^2)    \right. \nonumber \\
+   \left.   \frac{3}{70} (8B^4+3C^4+24 B^2C^2) \right]
- \frac{48}{25}  \left[
 \left(  \frac52 + 2 B+C  \right)^2+  C^2 - \frac54 ( \omega^2+1 )
\right]^2.
\label{q}
\end{align}

Before reducing the number of equations any further, it is fitting to note that 
 the system \eqref{A24}, \eqref{A25} and \eqref{A18} has a root 
reproducing the asymptotic expansion \eqref{A1}:
\be
B= - \frac{\epsilon^2}{4}, \quad C= \frac{\epsilon^2}{12}, \quad b^{-2}=  \epsilon^2.
\label{as1}
\ee
Here we have defined a small parameter $\epsilon$  by letting, in equations \eqref{A24}, \eqref{A25} and \eqref{A18},
  $\omega^2=4-  \epsilon^2$.
Using  \eqref{A13} we recover the amplitude of the outstanding (first) harmonic in the Dashen-Hasslacher-Neveu's expansion:
\be
\label{as2}
A^2= \frac{ \epsilon^2}{3}.
\ee

Returning to the system of three equations and using equation \eqref{A24} to eliminate $b^{-2}$ from \eqref{A25} and \eqref{A18},
we arrive at 
\begin{subequations}
\begin{align} 
\left( B+ \frac54 \right) f_\omega(B,C)  - \left( C + \frac{10}{4} \right) g_\omega(B,C)=0, 
\label{A41}   \\
f_\omega^2(B,C)  + \left( C+\frac{10}{4} \right) p_\omega(B,C)  f_\omega(B,C)  + \left( C+\frac{10}{4} \right)^2 q_\omega(B,C)=0.
\label{A42}
\end{align}
\label{sysBE}
\end{subequations}
For each $\omega$, equations \eqref{sysBE} 
with $f_\omega,   g_\omega, p_\omega,   q_\omega$ as in \eqref{f}, \eqref{g}, \eqref{p} and \eqref{q},  comprise a system of two equations 
with two unknowns, $B$ and $C$.

\begin{figure}[t]
 \begin{center} 
            \includegraphics*[width=0.49\linewidth] {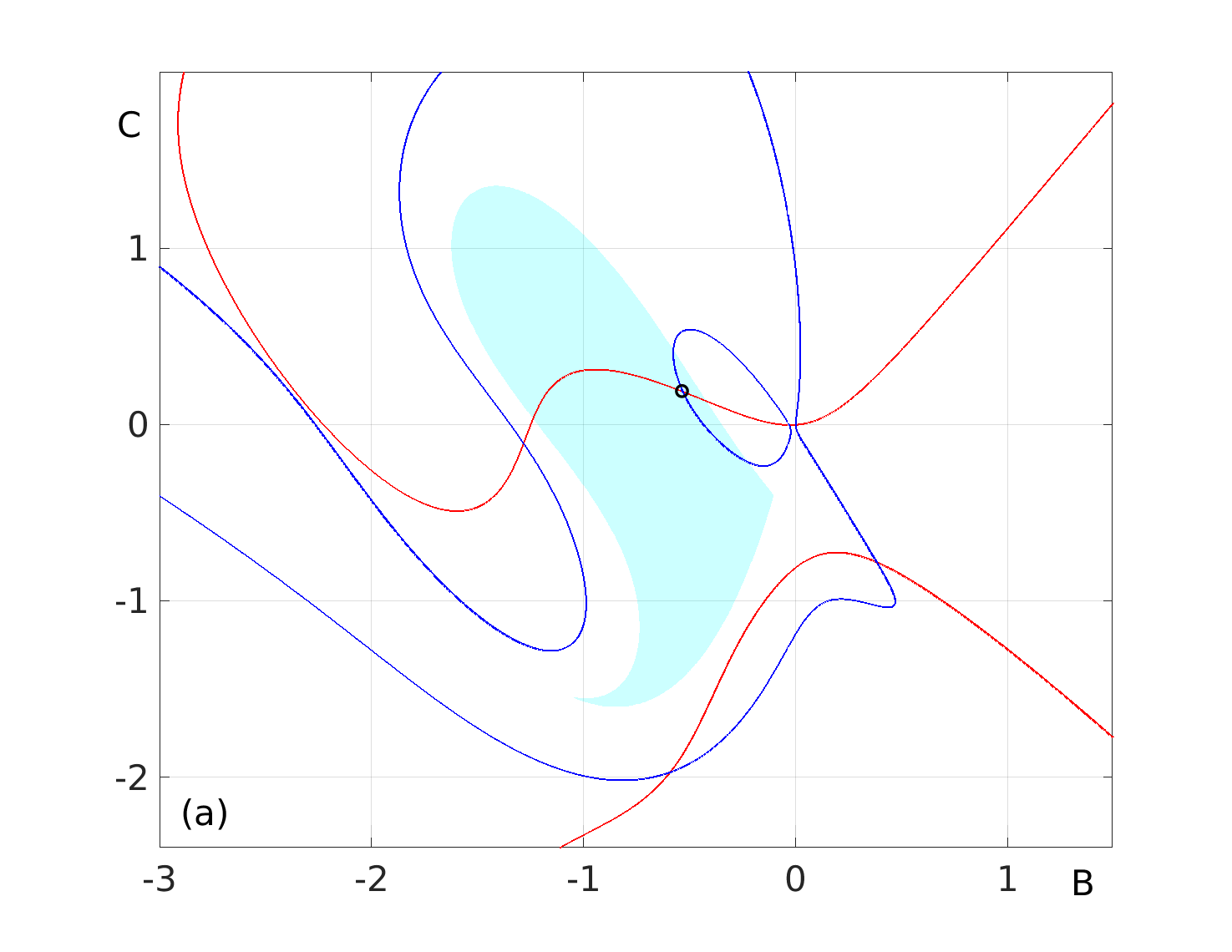}    
            \includegraphics*[width=0.49\linewidth] {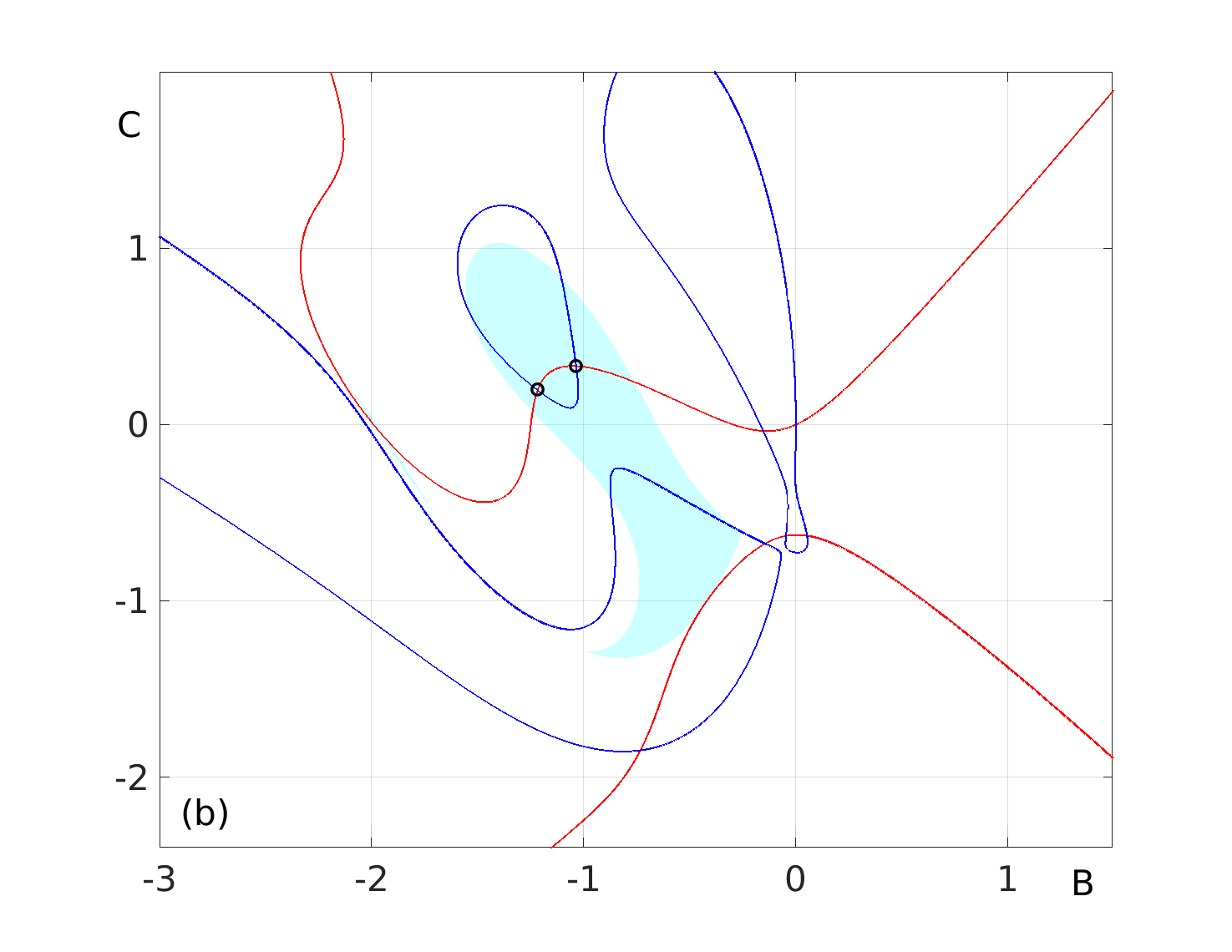}   
                                 \end{center}
  \caption{Graphical solution of the system  \eqref{sysBE} with $\omega=0.71 \, \omega_0$ (a) and $\omega=0.45 \, \omega_0$ (b).
   The red and blue curves  are described by equations \eqref{A41} and  \eqref{A42}, respectively.
     Tinted in light blue  is the \enquote{physical}  region where the inequalities $A^2>0$ and $b^{-2}>0$ are satisfied simultaneously,
  with $A^2$ and $b^{-2}$ as in \eqref{A80} and \eqref{A24}. Circles mark the roots of the system in the  physical  region. 
    \label{BE}  
  }
 \end{figure}

For much of the $\omega$, the system  \eqref{sysBE} has multiple  roots with real $B$ and $C$ (Fig \ref{BE}). 
However only roots satisfying $A^2>0$ and $b^{-2}>0$ correspond to fixed points of the variational equations \eqref{vareq}.
Here $b^{-2}$ is given by \eqref{A24} and $A^2$ by
  equation 
% \label{A80}
%  A^2= 1+ \omega^2- \frac{f_\omega(B,E)}{3(E+\frac52)} -\frac45 E^2-\frac45 \left( \frac52+ 2B+E \right)^2
  \begin{subequations} 
 \label{A80}  \be
A^2= \frac{h_\omega(B,C)}{2C+5}
\ee
with
\be
h_\omega= - 2B^2 - C^2 -B 
\left[ \omega^2+ \frac83 + \frac27 (2B^2+3C^2) -\frac45C^2 -\frac45\left( \frac52 + 2B+C\right)^2 \right],
\ee
\end{subequations}
which     ensues from \eqref{A13}-\eqref{A14}.

\begin{figure}[t]
 \begin{center} 
     \includegraphics*[width=0.60\linewidth] {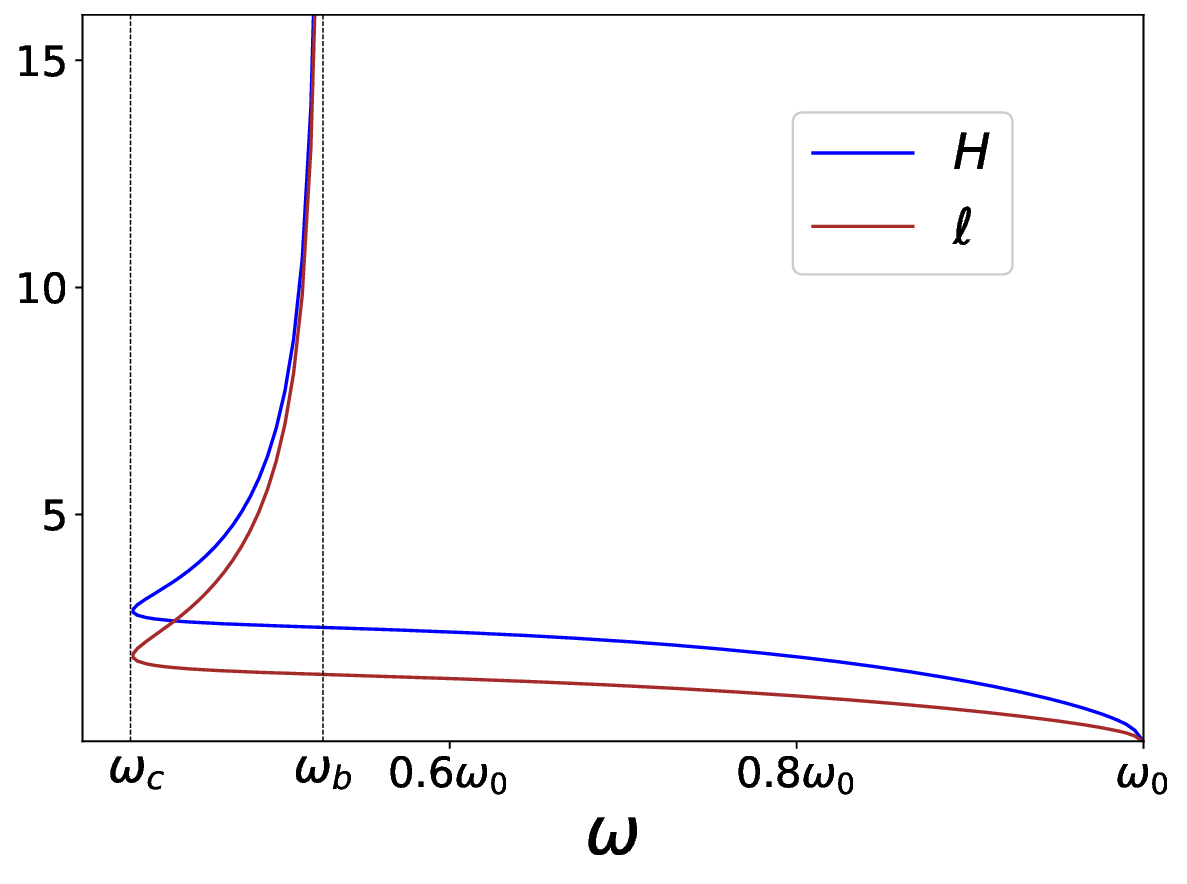}    
                                 \end{center}
  \caption{The adiabatic invariant $\ell$ and energy $H$ 
  of the fixed point as functions of $\omega$.  
    \label{Eom}  
  }
 \end{figure}

    There is only one such root when $\omega$ lies in the interval $(\omega_b, \omega_0)$, where
$\omega_b= 0.527 \, \omega_0$ (and $\omega_0=2$).
 In
the vicinity  of $\omega_0$, the corresponding fixed point is given by equations \eqref{as1}-\eqref{as2}. 
 The branch of fixed points extends from $\omega_0$  to 
  $\omega_c=0.417 \, \omega_0$ where it folds onto itself. As we path-follow  the turning branch back to 
   $\omega_b$, the $A$, $B$ and $C$ components  of the fixed point approach finite values while 
   $b$
  grows without bound. As a result, the adiabatic invariant $\ell=\omega b (A^2+\frac83 C^2)$ and 
  the  energy \eqref{Hen} tend to infinity as well (Fig \ref{Eom}).

 \section{Appendix: Linearisation matrix}
 \label{BBB} 
 
 The matrix elements $M_{ij}=M_{ji}$ in equation \eqref{MP} are given by
\begin{align}
	M_{11} =  24 \omega^2   \frac{  A^{2}  b}{3 A^{2}+8 C^{2}} +\frac{2}{3 b}+2b    \left[  3A^2+  \frac{4}{5} \left( \frac{5}{2}   +   2B+C       \right)^{2}+\frac{4 C^{2}}{5}-\omega^{2}-1\right]; 
	\nonumber \\
	 M_{12} = \frac{32}{5}    Ab \left(  \frac52  +2 B+ C \right); \quad
	M_{13}=      64 \omega^2 \frac{AC b }{3 A^{2}+8 C^{2}}+ \frac{16 }{5}   Ab \left(   \frac52  +2B+2 C\right);   \nonumber    \\	
	M_{14}=    4  \omega^{2}  A  + \frac{8 }{5} A   \left(\frac52 + 2B+C \right)^{2}    +2   A \left(A^{2}+\frac{4 C^{2}}{5}                   -\frac{1}{3 b^{2}}                         -\omega^{2}-1   \right);  \nonumber  \\
		M_{22}=\frac{32}{5} A^2  b +\frac{32}{15 b}+\frac{32}{105} b \left[36\left(B+\frac{7}{6}\right)^{2}+18 C^{2}-14\right]; \quad
	M_{23}  =  \frac{16 }{35} b \left(7 A^{2}+24 C B+28 C\right);  \nonumber    \\
	M_{24}=    \frac{16}{5} A^2 \left(  \frac{5}{2}+2B +C \right) + 
	\frac{192 }{35}    \left( B+ \frac{7}{6}\right) C^{2}+   \frac{8}{15}  B  \left[ \frac{48}{7}  \left( B+ \frac74 \right)^2-1 \right] 
-   \frac{32}{15} \frac{B}{b^{2}};  
	\nonumber   \\
	M_{33}=   \frac{512}{3}   \omega^{2}  \frac{C^{2}b }{3 A^{2}+8 C^{2}}                 
	 + \frac{192 }{35} b\left(B+\frac{7}{6}\right)^{2}+\frac{16}{15 b}
	 +\frac{16}{105}   b \left(21 A^{2}+27 C^{2}-35 \omega^{2}-14\right);   \nonumber    \\
	M_{34}=   \frac{32}{3}   \omega^{2} C+ \frac{48 }{35} C^3  +\frac{16}{15}    \left[       3 A^{2}     +  \frac{36}{7}    \left(B+\frac{7}{6}\right)^{2}-5 \omega^{2}-2\right]C+\frac{16}{5} A^2 \left(B+\frac{5}{4}\right)   -\frac{16}{15} \frac{C}{b^2};    \nonumber   \\
	M_{44}=      \frac23  \frac{\omega^2}{b} (3A^2+8C^2)
	 + \frac{2}{15 b^{3}} \left(5 A^{2}+16 B^{2}+8 C^{2}\right).
	\label{Ms} 
\end{align}
Here  $A, B, C$ and $b$ are components of the real root of the system \eqref{fp}.

Note that the linearisation of equations \eqref{vareq} was carried out under the assumption that 
 $\ell$ is a fixed, unperturbed, parameter: $\delta \ell=0$.
%  while  the frequency would be varied according to equation \eqref{ome}. 
Also note that the frequency $\omega$ does not appear in \eqref{vareq}.
However, 
once the linearisation procedure has been completed, we reintroduce $\omega$ according to 
	\[
\omega= \frac{\ell}{b \left( A^2 + \frac83 C^2 \right)},
% \label{ome}
\]
in agreement with  \eqref{ome}. 
In  the matrix elements \eqref{Ms}, the variables $A, B, C$ and $b$ are
single-valued functions of  $\ell$ --- and $\omega$ is also considered as a function of $\ell$.

\end{document}